\documentclass[prb,preprint,showpacs]{revtex4}
\begin{document}
\title{Martensitic Transition, Ferrimagnetism and Fermi Surface Nesting in Mn$_{2}$NiGa}
\author{S. R. BARMAN$^{1}$,  S. BANIK$^{1}$, A. K. SHUKLA$^{1}$, C. KAMAL$^2$, and APARNA CHAKRABARTI$^2$}
\affiliation{$^1$UGC-DAE Consortium for Scientific Research, Indore, 452017, Madhya Pradesh, India}
\affiliation{$^2$Raja Ramanna Centre for Advanced Technology, Indore, 452013, Madhya Pradesh, India} 

\begin{abstract}

\vskip 0.5cm
The electronic structure of  Mn$_2$NiGa has been studied using density functional theory and photoemission spectroscopy. The lower temperature tetragonal  martensitic phase with $c/a$=\,1.25 is more stable compared to the higher temperature austenitic phase. Mn$_2$NiGa is ferrimagnetic in both phases.  The calculated valence band spectrum, the optimized lattice constants and the magnetic moments are in good agreement with experiment. The majority-spin Fermi surface (FS) expands in the martensitic phase, while the minority-spin FS shrinks. FS nesting indicates occurrence of phonon softening and modulation in the martensitic phase. 
\end{abstract}

\maketitle

\noindent {\it{\bf Introduction:}} Recent advent of multiferroic shape memory alloys (SMA) like Ni-Co-Mn-In, Ni-Mn-Ga  that exhibit both ferroelastic and ferromagnetic properties has ushered a flurry of activity in this field\cite{Kainuma06,Takeuchi03,Sozinov02,Murray00,Marioni03,Biswas05,Shapiro07,Barman05,Banik06}.
In particular,   Ni-Mn-Ga has generated immense  interest 
~because of very large strain (10\%) in a moderate magnetic field ($\approx$1 Tesla)\cite{Sozinov02,Murray00}. Moreover, in Ni-Mn-Ga the actuation is much faster ($\approx$2\,kHz) than conventional SMA\cite{Marioni03}. 
However, Ni$_2$MnGa are brittle and so search for materials with better mechanical properties exhibiting similar magnetic field induced strain is being actively pursued\cite{Krenke05,Liu05}.  Mn$_2$NiGa is a recently discovered ferromagnetic SMA in the Ni-Mn-Ga family. It has Curie and martensitic start temperatures of 588 and 270~K, respectively\cite{Liu05}. 
Ferromagnetism in Mn$_2$NiGa 
~is surprising because direct Mn-Mn interaction normally leads to antiferromagnetic alignment\cite{Hobbs03,Zener51}. Moreover, the origin of the martensitic transition involving a relatively large tetragonal distortion ($c/a$=\,1.21) has not been studied theoretically till date. 
~Recently, a density functional theory (DFT) study on Mn$_2$NiGa shows a large enhancement of the density of states (DOS) near the Fermi level ($E_F$) and quenching of Mn and Ni magnetic moments in the martensitic phase\cite{Liu06}.  However, such large change in the magnetic moments or DOS has not been observed in any other SMA 
~either from experiment\cite{Banik06,Brown99,Banik07a} 
~or theory\cite{Barman05,Fuji89,Godlevsky01}.

~The geometry of the Fermi surface (FS) is responsible for a variety of phenomena  like spin or charge density waves, Kohn anomalies, Friedel oscillations in metals. If the FS has  parallel  planes, strong electronic response can occur at the wave vector that translates one parallel  plane of the FS to the other. This wave vector is called the nesting vector (n.v.).  FS nesting has been reported to cause softening of the transverse-acoustic (TA$_2$) phonon mode along [110] direction resulting in  modulated pre-martensitic  phase of SMA's like  Ni$_2$MnGa and Ni-Ti\cite{Bungaro03}. Recently, an inelastic neutron scattering study on Ni$_2$MnGa showed the presence of charge density wave  in the martensitic phase resulting from FS nesting\cite{Shapiro07}. Thus, it is worthwhile to study the FS of  Mn$_2$NiGa, particularly because the relatively large tetragonal distortion is likely to modify the FS substantially.

~In this work, a DFT study of the electronic structure of Mn$_2$NiGa using full potential linearized augmented plane wave method (FPLAPW) is presented. The valence band (VB) spectrum, calculated from the theoretical DOS, is in  agreement with the ultra-violet photoemission spectroscopy (UPS). We find that the total energy ($E_{tot}$) is lower in the martensitic phase with a tetragonal distortion of $c/a$=\,1.25.  We show that  Mn$_2$NiGa is an itinerant ferrimagnet in both the martensitic and austenitic phases. 
~The equilibrium lattice constants and the magnetic moments  are in  agreement with x-ray diffraction and magnetization data, respectively. 
~The FS in the martensitic phase is drastically different from the austenitic phase. A highly nested hole-type majority-spin cuboidal FS sheet around the $\Gamma$ point appears in the martensitic phase that is absent in the austenitic phase.

\noindent {\bf Methodology:}  
First principles DFT calculations were performed using the WIEN97 code\cite{Wien97}.  Generalized gradient approximation (GGA) for the exchange correlation that accounts for the density gradients was used\cite{Perdew96}. 
~An energy cut-off for the plane wave expansion of 16 Ry is used ($R_{MT}$$K_{max}$=~9).  The cut-off for charge density is $G_{max}$=~14. The maximum $l$ ($l_{max}$) for the radial expansion is 10, and for the non-spherical part: $l_{max,ns}$=\,6.  The muffin-tin radii are Ni: 2.1364, Mn: 2.2799, and  Ga: 2.1364~a.u. The number of $k$ points for self-consistent field cycles in the irreducible Brilloiun zone is 256 and 484 in the austenitic and martensitic phase, respectively. 
~The convergence criterion for $E_{tot}$ is  0.1 mRy, which implies that accuracy of $E_{tot}$  is $\pm$0.34 meV/atom.  The  charge convergence is set to 0.001. FS has been calculated using XcrySDen\cite{Kokalj99}.  Mn$_2$NiGa ingot was prepared by arc furnace melting and  annealing at 1100\,K\cite{Banik06}. It was characterized by x-ray diffraction (XRD), energy dispersive analysis of x-rays and differential scanning calorimetry\cite{Banik07a}.   Atomically clean specimen surface was prepared by {\it in situ}  scraping using a diamond file and the chamber base pressure was 6$\times$10$^{-11}$ mbar. UPS was performed with a He\,I (h$\nu$\,=21.2~eV) photon source using electron energy analyzer from Specs GmbH, Germany. The overall resolution was 120\,meV. 


Mn$_2$NiGa  has a cubic $L_{2_1}$ structure in the austenitic phase that consists of four interpenetrating f.c.c. lattices at (0,\,0,\,0), (0.25,\,0.25,\,0.25), (0.5,\,0.5,\,0.5), and (0.75,\,0.75,\,0.75) (Fig.~1a)\cite{Liu05,Banik07a}. The structure of Mn$_2$NiGa can be better explained in comparison to Ni$_2$MnGa that also has  $L_{2_1}$ structure. 
~In Ni$_2$MnGa, the Ni atoms are at (0.25,\,0.25,\,0.25) and (0.75,\,0.75,\,0.75), while Mn and Ga are at (0.5,\,0.5,\,0.5) and (0,\,0,\,0), respectively and there is no direct Mn-Mn interaction, with Mn having eight Ni atoms as nearest neighbours. In contrast, Mn$_2$NiGa has one Mn atom at (0.5,\,0.5,\,0.5) (referred to as MnII), while the other Mn atom (MnI) occupies the Ni atom position (0.75,\,0.75,\,0.75) of Ni$_2$MnGa. Thus, MnI and MnII occupy inequivalent sites in the unit cell, and there is a direct Mn-Mn interaction since MnI and MnII are nearest neighbours. In the martensitic phase, the XRD pattern for  Mn$_2$NiGa has been indexed by a tetragonal unit cell with $c/a$=\,1.21 (Fig.~1b)\cite{Liu05,Banik07a}.

\noindent{\bf Total energy  and magnetic moment calculation:} To determine whether minimization of $E_{tot}$ causes the structural transition, we have calculated $E_{tot}$ for both phases as a function of the lattice parameters in the lowest energy magnetic state (discussion about the magnetic state is given later).  In the austenitic phase, $E_{tot}$ as a function of cell volume ($V$)  exhibits a parabolic behaviour and the minimum (shown by arrow) determines the optimized lattice constant ($a$=\,11.059 a.u.=\,5.85\,\AA)~(Fig.~2a). 
 ~The agreement is within 1\% of the experimental value of 5.9072\,\AA\cite{Liu05}.
~For the martensitic phase, in the first step, $E_{tot}(V)$ is calculated to obtain optimized $V$=\,1330~a.u.$^3$ at fixed $c/a$=\,1.21 (XRD value).  
~ Next, $E_{tot}$($c/a$) is calculated at  $V$=\,1330~a.u.$^3$. This gives the optimized  $c/a$ to be 1.25. In the final step, $E_{tot}$($V$) is calculated again with $c/a$=\,1.25 (Fig.~2a). Least square fitting of the data\cite{Barman05,Chakrabarti01} gives the $E_{tot}$ minimum at 1335.2~a.u.$^3$ (shown by arrow). 
~From Fig.~2a, the $E_{tot}$ minimum in the martensitic phase is 6.8 meV/atom lower than the austenitic phase.  
~This demonstrates that the martensitic phase is stabilized through a sizable tetragonal distortion  ($c/a$=\,1.25).
~The optimized lattice constants ($a$=\,5.409 and $c$=\,6.762, \AA)~
 ~are within 2.1\% and 0.85\% of the experimental lattice constants $a$=\,5.5272\,\AA~ and $c$=\,6.7044\,\AA, respectively\cite{Liu05}.  Thus, the agreement of the lattice constants for both the phases is satisfying, considering that even for free-electron-like non-magnetic metals there could be about 2\% discrepancy between experiment and GGA based DFT theory\cite{Perdue92}. 
The decrease of $V$  by 1.2\% is in agreement with the experimental volume decrease of 0.64\% in the martensitic phase\cite{Liu05}.



The lowest energy magnetic state is obtained by performing $E_{tot}$ minimization over various possible starting MnI and MnII magnetic moment combinations, as  discussed in details in Ref.\cite{Barman07a}. For both austenitic and martensitic phase, the  anti-parallel starting spin (equal or unequal) configurations  of MnI and MnII converge to a  ferrimagnetic state that has minimum $E_{tot}$. We have used starting Mn magnetic moments for structure optimization runs to be 3$\mu_B$ for both Mn atoms in anti-parallel orientation. However, when the starting MnI and MnII moments are  parallel (equal or unequal), $E_{tot}$ converges to different magnetic moments related to local minima at higher energies. For example, in the austenitic phase there are three local minima\cite{Barman07a}. 
~Also in the martensitic phase,  multiple local minima are obtained with parallel starting moments of MnI and MnII. In particular, a local minimum that is 108 meV/atom higher in $E_{tot}$, gives  MnI and  MnII moments to be  0.24 and 2.38 $\mu_B$\cite{Barman07a}. Thus, one Mn moment is  small, as has been reported in Ref.\cite{Liu06}. Our calculation based on the magnetic moments reported in Ref.\cite{Liu06} converges  at 193 meV/atom higher energy than the $E_{tot}$ minimum\cite{Barman07a}. 
~This gives an idea why the results from Ref.\cite{Liu06} are in disagreement with experimental data, as discussed later.

The spin magnetic moment distribution in the martensitic phase  clearly shows that it is ferrimagnetic with MnI magnetic moment anti-parallel and smaller than MnII (Fig.~2b). Ni moment is small and is parallel to MnII moment.  
~For the martensitic (austenitic) phase, the local spin magnetic moments 
~are -2.21~(-2.43), 2.91~(3.2), 0.27~(0.32), 0.01 (0.01) $\mu_B$ per formula unit ($\mu_B$/f.u.) for MnI, MnII, Ni, and Ga, respectively.  The moment related to the interstitial charge is small (-0.04\,$\mu_B$). 
~The total moment for the martensitic phase (1.01\,$\mu_B$/f.u.) is 11\% less than the austenitc phase (1.14\,$\mu_B$/f.u). The lowering of the magnetic moment in the martensitic phase has been reported by Liu {\it et al.} from magnetization studies\,: 1.21\,$\mu_B$/f.u. (28.28 emu/g) and 1.29\,$\mu_B$/f.u. (30.3 emu/g) in the martensitic and austenitic phase, respectively\cite{Liu05}. Thus, the magnetic moment values and the trend that magnetization is lower in the martensitic phase are in agreement with our calculations. 


\noindent{\bf Density of states and photoemission spectroscopy:} The stabilization of the tetragonally distorted martensitic phase in Ni$_2$MnGa has been related to band Jahn-Teller effect, where a DOS peak at $E_F$ in the cubic phase splits into two peaks below and above $E_F$ in the tetragonal phase, resulting in a lowering of the total energy\cite{Fuji89}. Splitting and shift of the DOS peaks just below $E_F$ have also been observed in    Ni$_{2.25}$Mn$_{0.75}$Ga\cite{Banik06}. For Mn$_2$NiGa, the differences in the total DOS near $E_F$ are interesting: a peak at -0.1~eV  in the austenitic phase shifts to lower energy (-0.35~eV) and diminishes in intensity in the martensitic phase (both peaks indicated by arrows).  The peak above $E_F$ at 0.35~eV (tick) does not shift but is enhanced in intensity  in the martensitic phase indicating a transfer of DOS from the occupied to the unoccupied states.  From the partial DOS (PDOS), it is clear that the peaks at -0.1 and  -0.35~eV arise primarily due to Ni 3$d$ and MnI 3$d$ hybridization. The shift of the -0.1~eV peak to lower energy in the martensitic phase results from enhanced  Ni 3$d$- MnI 3$d$ hybridization caused by decrease in Ni-MnI distance from 2.925\,\AA~(austenitic) to 2.701\,\AA~(martensitic) and is a possible reason for the stabilization of the martensitic phase.   The DOS at $E_F$ is substantially reduced  in the martensitic phase (1.29 states/eV\,f.u.) compared to the austenitic phase (3.39). Thus,  decrease in electronic specific heat in the martensitic phase could be expected.


The antiferromagnetic alignment of MnI and MnII spin moments can be understood from the 3$d$ spin resolved PDOS (Fig.~3b). MnI 3$d$ minority-spin states appear below $E_F$ between -1 to -3.5\,eV, whereas MnII 3$d$ majority-spin states appear below $E_F$ with two well separated high PDOS region around -1.5 and -2.7\,eV. MnI 3$d$ majority-spin states appear primarily above $E_F$ centered around  0.7\,eV; while MnII 3$d$ minority-spin states appear above $E_F$ with the main peak at 1.1\,eV and a smaller peak at 0.35\,eV.   Thus, while the minority-spin states are mostly excluded from the MnII 3$d$ shell, the majority-spin states are excluded from the MnI 3$d$ shell resulting in large but oppositely aligned moments. MnI and MnII are nearest neighbors (n.n.) with n.n. distance of  2.549 (2.533)\,\AA~ in the martensitic (austenitic) phase.  The exchange pair interaction as a function of Mn-Mn separation  was calculated by a Heisenberg-like model and an antiferromagnetic coupling at short interatomic distances was  found that becomes ferromagnetic at larger distances\cite{Hobbs03}. Thus, direct Mn-Mn interaction at  short interatomic distance is responsible for their opposite alignment\cite{Hobbs03,Zener51}. The energy separation between the centroid of the occupied  and the unoccupied  spin states of opposite polarization gives an exchange splitting of   2.7\,eV (3.1\,eV) for MnI (MnII) in the martensitic phase. In the austenitic phase, the exchange splittings are  2.8 and 3.6~eV for MnI and MnII, respectively. Thus, the Stoner parameter (ratio of exchange splitting and magnetic moment) is roughly  about 1 eV/$\mu_B$ in both phases, which is characteristic of itinerant magnetism.

It was shown for Mn excess Ni$_2$Mn$_{1+x}$Ga$_{1-x}$ that the magnetic moments of Mn atom in Ga site is equal but anti-parallel to the Mn atom at Mn site\cite{Enkovaara03}. This would tend to suggest that in Mn$_2$NiGa, the Mn moments would cancel and a small total moment might result from Ni. However, this does not happen and the difference of MnI and MnII moments is key to the larger total moment ($\approx$1\,$\mu_B$). This originates from the stronger hybridization  between the majority-spin  Ni and MnII 3$d$ states in comparison to hybridization between  Ni and MnI 3$d$ minority-spin states. Note that Ni and MnII are n.n. separated by 2.549 (2.533)\,\AA~ in the martensitic (austenitic) phase and stronger hybridization pulls down almost all the MnII 3$d$ majority-spin states below $E_F$ resulting in strong spin polarization and larger moment.  On the contrary, hybridization between  Ni and MnI 3$d$ minority-spin states is relatively weaker, distance being  larger: 2.701 (2.925)\,\AA~ in the martensitic (austenitic) phase, and there are sizable  MnI 3$d$ minority-spin states above $E_F$ including the 0.35~eV peak, resulting in smaller moment on MnI.


Photoemission spectroscopy is a direct probe of the DOS in the VB region. In Fig.~4, the main peak of the UPS VB spectrum appears at -1.4~eV and the Fermi cut-off is at 0~eV. In order to calculate the VB spectrum, we note because of the order of magnitude larger photoemission cross-sections of Ni 3$d$ and Mn 3$d$ (4.0 and 5.3 mega barns at  h$\nu$=\,21.2 eV, respectively)\cite{Lindau}, these PDOS determine the shape of VB\cite{Chakrabarti05}. So, we have added the Ni and Mn 3$d$  PDOS in proportion to their cross-sections,  multiplied by the Fermi function and broadened by  the instrumental Gaussian resolution and the life-time width related energy dependent Lorenzian to obtain the calculated VB (Fig.~4). This is a standard procedure of comparing the photoemission spectrum from a polycrystalline sample with the calculated DOS\cite{Chakrabarti05,Sarma95}. 
 ~The position of the main peak at -1.4~eV  and the ratio between the main peak and the intensity at $E_F$ are in good agreement with UPS VB spectrum. 
 ~It is clear from Fig.~4 that the main peak is dominated  by Mn 3$d$\,-\,Ni 3$d$ hybridized states that have almost equal contribution. States near $E_F$ are dominated by Mn 3$d$ states, and the MnI 3$d$ in particular. 
 
The martensitic phase DOS from Ref.~\cite{Liu06}, obtained by adding up the majority and minority-spin DOS from Fig.~5 of Ref.\cite{Liu06}, is in clear disagreement with our DOS (Fig.~3a). This prompted us to calculate  the VB spectrum from the PDOS of Ref.\cite{Liu06} following the same procedure as discussed above and compare it with the experimental UPS VB. As shown in Fig.~4, the calculated VB based on Ref.\cite{Liu06} is  in obvious disagreement with UPS VB: no clear peak is observed in the former;  a weak broad feature is present at -2~eV and the intensity near $E_F$ is highest. This shows that the martensitic phase DOS reported in Ref.\cite{Liu06} is inconsistent with experiment.    Moreover, the large change of local moments (austenitic MnI=\,-2.2, MnII=\,3.15, Ni 0.27 $\mu_B$ to martensitic MnII=\,Ni\,$\approx$0, MnI=\,1.4 $\mu_B$) obtained in Ref.\cite{Liu06} is physically unexpected\cite{Barman07a}, since the MnI\,-\,MnII distance change by only 0.6\% in the martensitic phase. Thus, it is no wonder why the total moment reported in Ref.\cite{Liu06} is higher in the martensitic phase compared to the austenitic phase, in contradiction to their own magnetization data\cite{Liu05,Liu06}.  


\noindent {\bf Electronic bands and Fermi surface:} {\it Austenitic phase majority spin states:} We now turn to the discussion of the electronic bands and Fermi surface of Mn$_2$NiGa. The majority-spin bands in the austenitic phase show that band 29 forms electron pockets (Fig.~5b). The corresponding FS,  shown in Fig.~5d, is distorted prolate ellipsoidal in shape and occurs around the $X$ point of the Brillouin zone (BZ) with the long axis along the $\Gamma$$X$ direction.  The BZ is shown in Fig.~5a. The projection of the FS along $\Gamma$$X$ is a square (inset, Fig.~5d), which indicates that the FS nests onto itself with n.v. 0.44(1,0,0) and 0.44(0,1,0), in units of 2$\pi$/$a$ (=\,1 a.u.).  
The nested portion of the FS is a rhombus (shown by black lines in Fig.~5d) of area 0.052 a.u.$^2$ with an opening angle of about 15$^{\circ}$.  


\noindent {\it Martensitic phase majority spin states:} In the martensitic phase, the majority-spin FS exhibits interesting modification (Fig.~5e). The majority-spin band 29 related electron type FS is now connected as continuous pipes along (1,0,0) direction, but with varying cross-section  with flat parallel parts that nest onto each other (green/pink sheet in Fig.~5e). The n.v. are 0.34(1,0,0) and 0.34 (0,1,0), and compared to the austenitic phase the direction is same but  the magnitude of the n.v.'s is reduced. 
~Interestingly, a second majority-spin band (28) crosses $E_F$ that results in a hole-type cuboid FS around the $\Gamma$ point that has no counterpart in the austenitic phase (blue sheet, Fig.~5e). Two mutually perpendicular n.v.'s 0.75(1,1,0) and 0.75(1,-1,0) are identified, along with a larger n.v. of 1.13(0,0,1). The n.v.'s along the $\left\{1,0,0\right\}$, identified above, are not expected to contribute to phonon softening because these hardly contribute to the electron-phonon coupling matrix element\cite{Bungaro03}. 
On the other hand, the 0.75(1,1,0) and 0.75(1,-1,0) n.v.'s might be responsible for the softening of the TA$_2$[110] phonon resulting in a modulated martensitic phase.  The different nesting vectors are shown in Table I. 

\begin{table*}
\caption{\footnotesize Nesting vectors for the Fermi surface of Mn$_2$NiGa, in units of 2$\pi$/$a$ (=\,1 a.u.).}
\centering
\begin{tabular}{|p{1.5cm}|p{2.5cm}|p{2.5cm}|p{2.5cm}|p{2cm}|} \hline
\textbf{}  & \multicolumn{2}{|c|}{\text{Austenitic phase}}& \multicolumn{2}{|c|}{\text{Martensitic phase}}\\\hline
{Band no.} & {Majority spin} & {Minority spin} & {Majority spin} & {Minority spin}\\\hline
29  & 0.44(1,0,0), 0.44(0,1,0)  & -- & 0.34(1,0,0), 0.34(0,1,0) & --\\\hline 
28  & --  & 0.31\verb+{+1,0,0\verb+}+ &  0.75(1,1,0), 0.75(1,-1,0), 1.13(0,0,1) & -- \\\hline 
27  & --  & 0.4\verb+{+1,0,0\verb+}+ &  -- & -- \\\hline 
\end{tabular}
\end{table*}

 From Fig.~5d and e, the majority spin FS is clearly enlarged in the martensitic phase compared to the austenitic phase.
~ In the contrary, for the minority-spin states (Fig.~5f-i), the FS clearly shrinks in the martensitic phase. 

\noindent {\it Austenitic phase minority spin states:} In the austenitic phase, minority spin band 27 is hole-type dispersing above (below) $E_F$ at 0.2$\Gamma$$L$ (0.5$LW$) and generates distorted cubic FS, where one pair of diagonally opposite corners taper out 
~(Fig.~5f). FS nesting is observed between the cube faces with n.v. 0.4\verb+{+1,0,0\verb+}+,  as shown by the yellow arrows. 
~The second sheet of the FS (band 28) is electron-like, consisting of multiply connected pipes of square cross-section (inset, Fig.~5h). The parallel surfaces of the pipes nest onto each other with a n.v. of 0.31\verb+{+1,0,0\verb+}+\,a.u. and a nesting area of 0.16\,a.u.$^2$ 

\noindent {\it Martensitic phase minority spin states:} In the martensitic phase, the minority spin hole type FS (band 27) has a flower-like shape with a perforation in the middle (Fig.~5g). 
~The electron type FS sheet shrink to disconnected pipes of varying diameter (Fig.~5i).
~These minority-spin FS sheets (Fig.~5g,i) in the martensitic phase do not exhibit  nesting. 

\noindent{\bf Conclusion:} We observe FS nesting in the martensitic phase  along [1,1,0] direction in the majority-spin FS that might lead to the  instability of the TA$_2$ phonon mode in Mn$_2$NiGa.  The austenitic phase FS is drastically modified in the martensitic phase. The majority spin FS expands in the martensitic phase, while the minority-spin FS shrinks. 
~We show that Mn$_2$NiGa is an itinerant ferrimagnet in both austenitic and martensitic phase, and that the MnII or Ni moments do not become zero in the martensitic phase, refuting a recent work by Liu {\it et al.}\cite{Liu06}. ~The unequal spin magnetic moments in the two inequivalent Mn atoms  (MnI and MnII) arise from the difference in the hybridization  of the MnI~3$d$\,-\,Ni 3$d$ and MnII~3$d$\,-\,Ni 3$d$ states, which in turn is  related to the interatomic distances. We furthermore show that in Mn$_2$NiGa a large tetragonal distortion ($c/a$=\,1.25)  decreases the total energy, stabilizing the lower temperature martensitic phase. 
 ~Mn$_2$NiGa would be an ideal system to study different models of magnetization in metals since it has a simple $L_{2_1}$ structure and three sublattice magnetization with parallel (between MnII and Ni) and anti-parallel (between MnI and MnII) magnetic moment alignment. Possibility of incommensurate magnetic phase or charge density wave instabilities could be expected at  low temperatures due to presence of FS nesting and ferrimagnetism.   Low temperature x-ray diffraction might be able to detect possible occurrence of a charge density wave state.  Neutron scattering,  angle resolved photoemission or Compton scattering experiments can verify the theoretically predicted FS.   In fact, FS nesting, ferrimagnetism and large magnetoelastic coupling makes Mn$_2$NiGa a highly interesting material that has remained largely unexplored  so far.

We thank K. KUNC and A. DE SARKAR for fruitful discussions.  P. CHADDAH, V. C. SAHNI, K. HORN, A. GUPTA and S. M. OAK  are thanked for  support.  Ramanna Fellowship Research Grant and D.S.T.-Max Planck Partner Group Project are thanked for funding.


\newpage
\noindent {\bf Figure Captions}

\noindent {\bf Fig.~ 1} The structure of Mn$_2$NiGa in the (a) austenitic and (b) martensitic phase; the blue, green, red, and brown spheres represent Ni, MnI, MnII and  Ga, respectively.

\noindent {\bf Fig.~ 2} (a) The calculated total energies ($E_{tot}$) of Mn$_2$NiGa as a function of cell volume of the austenitic and martensitic phase. (b) Three dimensional plot of the spin magnetic moment  distribution (in unit of e\AA$^{-3}$) in the (110) plane in the martensitic phase, a contour plot is shown in the bottom.

\noindent {\bf Fig.~ 3} (a) Comparison of total density of states (DOS) and Ni 3$d$ and Mn 3$d$ partial DOS of Mn$_2$NiGa between the martensitic and austenitic  phases (b) minority- and majority-spin components of the DOS in the martensitic phase.

\noindent {\bf Fig.~ 4} UPS valence band (VB) spectrum of Mn$_2$NiGa in the martensitic phase compared with theoretical VB spectrum calculated from the DOS in Fig.~3a. The contributions from the Mn 3$d$ and the Ni 3$d$ partial DOS are also shown. The spectra have been shifted along the vertical axis for clarity of presentation.

\noindent {\bf Fig.~ 5} (a) The f.c.c. Brillouin zone showing the high symmetry directions. (b)  Majority and (c) minority-spin energy bands of Mn$_2$NiGa in the austenitic phase. Majority-spin Fermi surface (FS) of the (d) austenitic phase compared to the (e) martensitic phase FS related to bands 28 and 29. Minority-spin austenitic phase FS related to (f) band 27 and (h) band 28. Insets show the FS in a different orientation.   
~Martensitic phase minority-spin FS related to (g) band 27 and (i) band 28. All the FS are shown in the repeated zone scheme and yellow arrows represent the nesting vectors. Black arrows relate the FS of the two phases.

\end{document}